\documentclass[%
reprint,
superscriptaddress,
 amsmath,amssy
 aps,
prb,
]{revtex4-2}

\usepackage{graphicx}
\usepackage{bm}
\usepackage{ulem}
\usepackage{color}
\usepackage{mathrsfs}
\usepackage{gensymb}
\usepackage{upgreek}
\usepackage[colorlinks=true,linkcolor=blue,citecolor=blue]{hyperref}

\begin{document}

\title{Significant Unconventional Anomalous Hall Effect in Heavy Metal/Antiferromagnetic Insulator Heterostructures}
	
	\author{Yuhan Liang}
	\thanks{These authors contributed equally}
	\affiliation{School of Materials Science and Engineering, Tsinghua University, Beijing, 100084, China}
	
	\author{Liang Wu}
	\thanks{These authors contributed equally}
	\email{liangwu@kust.edu.cn}
	\affiliation{Faculty of Materials Science and Engineering, Kunming University of Science and Technology, Kunming, 650093, Yunnan, China}

	\author{Minyi Dai}
	\affiliation{Department of Materials Science and Engineering, University of Wisconsin-Madison, Madison, WI, USA}
		
	\author{Yujun Zhang}
	\affiliation{Institute of High Energy Physics, Chinese Academy of Sciences, Beijing 100049, China}
	
	\author{Qinghua Zhang}
	\affiliation{Institute of Physics, Chinese Academy of Sciences, Beijing 100049, China}
	
	\author{Jie Wang}
	\affiliation{Faculty of Materials Science and Engineering, Kunming University of Science and Technology, Kunming, 650093, Yunnan, China}
	
	\author{Nian Zhang}
	\affiliation{State Key Laboratory of Functional Materials for Informatics, Shanghai Institute of Microsystem and Information Technology, Chinese Academy of Sciences, Shanghai 200050, China}
	\affiliation{CAS Center for Excellence in Superconducting Electronics(CENSE), Chinese Academy of Sciences, Shanghai 200050, China}
	
	\author{Wei Xu}
	\affiliation{Institute of High Energy Physics, Chinese Academy of Sciences, Beijing 100049, China}
	
	\author{Le Zhao}
	\affiliation{Department of Physics, Tsinghua University, Beijing 100084, China}
	
	\author{Hetian Chen}
	\affiliation{School of Materials Science and Engineering, Tsinghua University, Beijing, 100084, China}
	
	\author{Ji Ma}
	\affiliation{Faculty of Materials Science and Engineering, Kunming University of Science and Technology, Kunming, 650093, Yunnan, China}
	
	\author{Jialu Wu}
	\affiliation{School of Materials Science and Engineering, Tsinghua University, Beijing, 100084, China}
	
	\author{Yanwei Cao}
	\affiliation{Ningbo Institute of Materials Technology and Engineering, Chinese Academy of Sciences, Ningbo 315201, China}
	\affiliation{Center of Materials Science and Optoelectronics Engineering, University of Chinese Academy of Sciences, Beijing 100049, China}
	
	\author{Di Yi}
	\affiliation{School of Materials Science and Engineering, Tsinghua University, Beijing, 100084, China}
	
	\author{Jing Ma}
	\affiliation{School of Materials Science and Engineering, Tsinghua University, Beijing, 100084, China}

	\author{Wanjun Jiang}
	\affiliation{Department of Physics, Tsinghua University, Beijing 100084, China}
	
	\author{Jia-Mian Hu}
	\affiliation{Department of Materials Science and Engineering, University of Wisconsin-Madison, Madison, WI, USA}

	\author{Ce-Wen Nan}
	\email{cwnan@tsinghua.edu.cn}
	\affiliation{School of Materials Science and Engineering, Tsinghua University, Beijing, 100084, China}
	
	\author{Yuan-Hua Lin}
	\email{linyh@tsinghua.edu.cn}
	\affiliation{School of Materials Science and Engineering, Tsinghua University, Beijing, 100084, China}

\date{\today}

\begin{abstract}
The anomalous Hall effect (AHE) is a quantum coherent transport phenomenon that conventionally vanishes at elevated temperatures because of thermal dephasing. Therefore, it is puzzling that the AHE can survive in heavy metal (HM)/antiferromagnetic (AFM) insulator (AFMI) heterostructures at high temperatures yet disappears at low temperatures. In this paper, we report that an unconventional high-temperature AHE in HM/AFMI is observed only around the N\'eel temperature of AFM, with large anomalous Hall resistivity up to 40 n$\Omega$ cm. This mechanism is attributed to the emergence of a noncollinear AFM spin texture with a non-zero net topological charge. Atomistic spin dynamics simulation shows that such a unique spin texture can be stabilized by the subtle interplay among the collinear AFM exchange coupling, interfacial Dyzaloshinski-Moriya interaction, thermal fluctuation, and bias magnetic field.
\end{abstract}

\maketitle
\section{Introduction}

Heavy metal (HM)/antiferromagnetic (AFM) insulator (AFMI) heterostructures are an emerging essential system for investigating the interaction between spin current and antiferromagnetic (AFM) order, which have the potential for applications in energy-efficient, ultrafast, and robust spintronics devices.\cite{Cheng2014, Fischer2018, Ji2018, Cheng2019, Moriyama2020,Han2020, Vaidya2020, Li2020, Jungwirth2016, Baltz2018} In particular, the reflected spin current after interactions with the magnetic order at the interface can provide valuable information for determining the magnetic order, which has already been well established in the counterpart HM/ferromagnet (FM) heterostructures, such as the spin Hall magnetoresistance (SMR) and its derivative spin Hall-anomalous Hall effect (SH--AHE).\cite{Nakayama2013, Chen2013} Most recently, considerable attention has been paid to SMR in HM/AFMI heterostructures, which is considered a probe for current-induced switching of the AFM order.\cite{Fischer2018, Manchon2017, Geprags2020, Moriyama2018, Chen2018} Despite the extensive investigations of SMR in HM/AFMI heterostructures, the anomalous Hall effect (AHE) in such systems has not yet been subjected to the same examination, therefore, its origin, particularly whether it is caused by SMR as in HM/FM heterostructures, remains unclear.

\begin{figure*}[th]
\centering
	\includegraphics[width=0.7\linewidth]{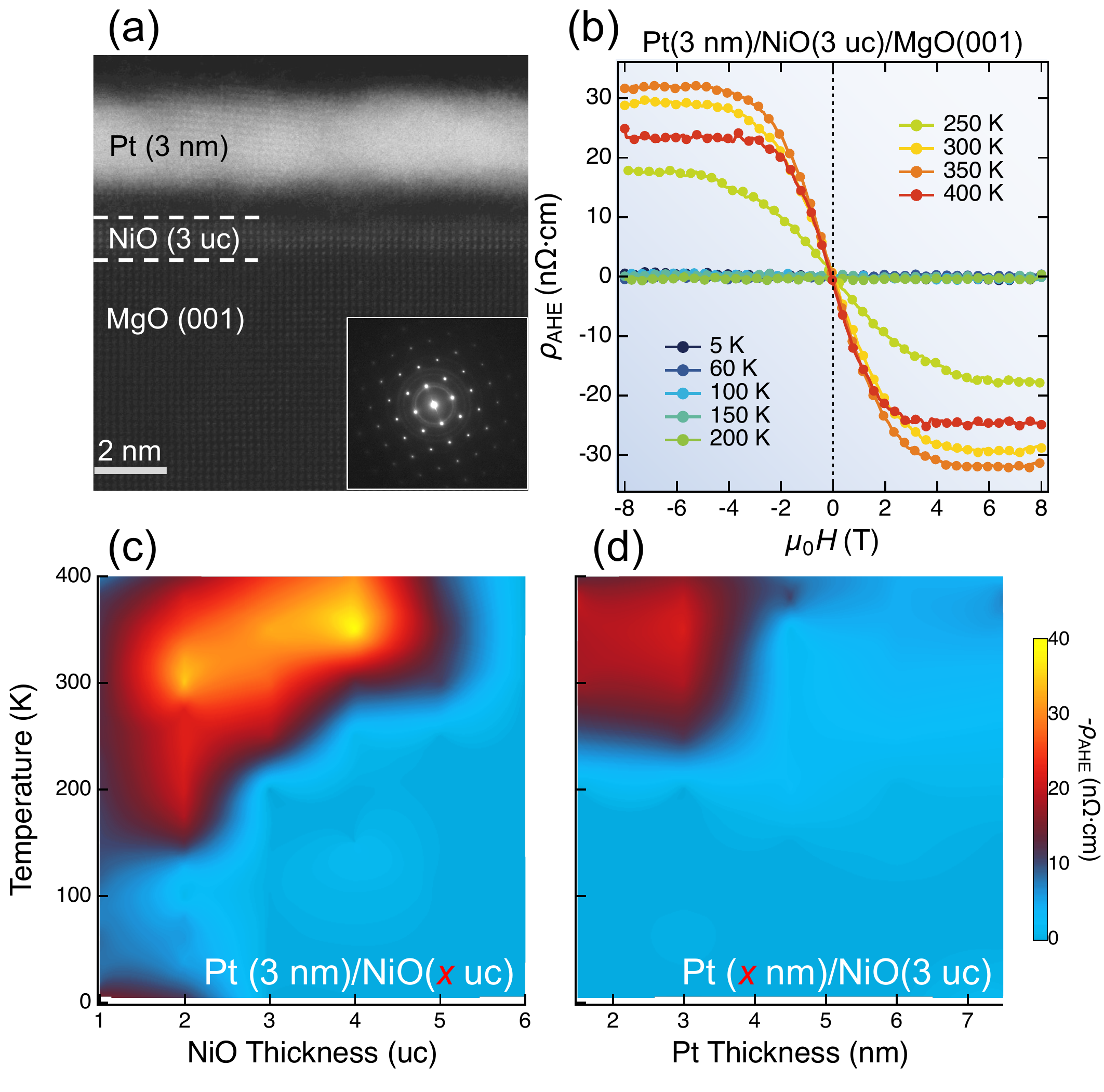}
	\caption{Observation of AHE in Pt/NiO/MgO heterostructures. (a) The STEM of Pt(3 nm)/NiO(3 uc)/MgO(001) (the inset is the SAED pattern). (b) $\rho_{\mathrm{AHE}}$ of the same sample at varying temperatures. (c-d) Evolution of the $\rho_{\mathrm{AHE}}$ with temperature and thickness of NiO and Pt.}
	\label{Fig1}
\end{figure*}

Recently, a high-temperature AHE was observed in HM/AFMI heterostructures, which are based on Ta, Pt, and W grown on Cr$_2$O$_3$ (AFMI) with an anomalous Hall resistivity ($\rho_\mathrm{AHE}$) of approximately 1 n$\Omega $ cm.\cite{Ji2018, Cheng2019, Moriyama2020} Despite the controversy regarding the underlying mechanism, a ubiquitous phenomenon insensitive to the selection of the HM and crystalline orientation of AFMI was observed. That is, a superparamagnetism-like AHE signal with a zero-coercive field ($H_\mathrm{c}$) exists at temperatures significantly higher than the bulk Cr$_2$O$_3$ N\'eel temperature $T_\mathrm{N}$ (307 K) but disappears in the low-temperature region ($<$ 200 K).\cite{Ji2018, Cheng2019, Moriyama2020} However, for heterostructures consisting of AFMI with significantly higher bulk $T_\mathrm{N}$, for example, the $\alpha$-Fe$_2$O$_3$ ($T_\mathrm{N} \sim$ 950 K) and NiO ($T_\mathrm{N} \sim$ 523 K), an AHE at approximately room temperature has not been reported. Thus, it is desirable to extend the AHE to other AFMI-based heterostructures, and understand its physical mechanism.

\begin{figure*}[thb]
\centering
	\includegraphics[width=0.7\linewidth]{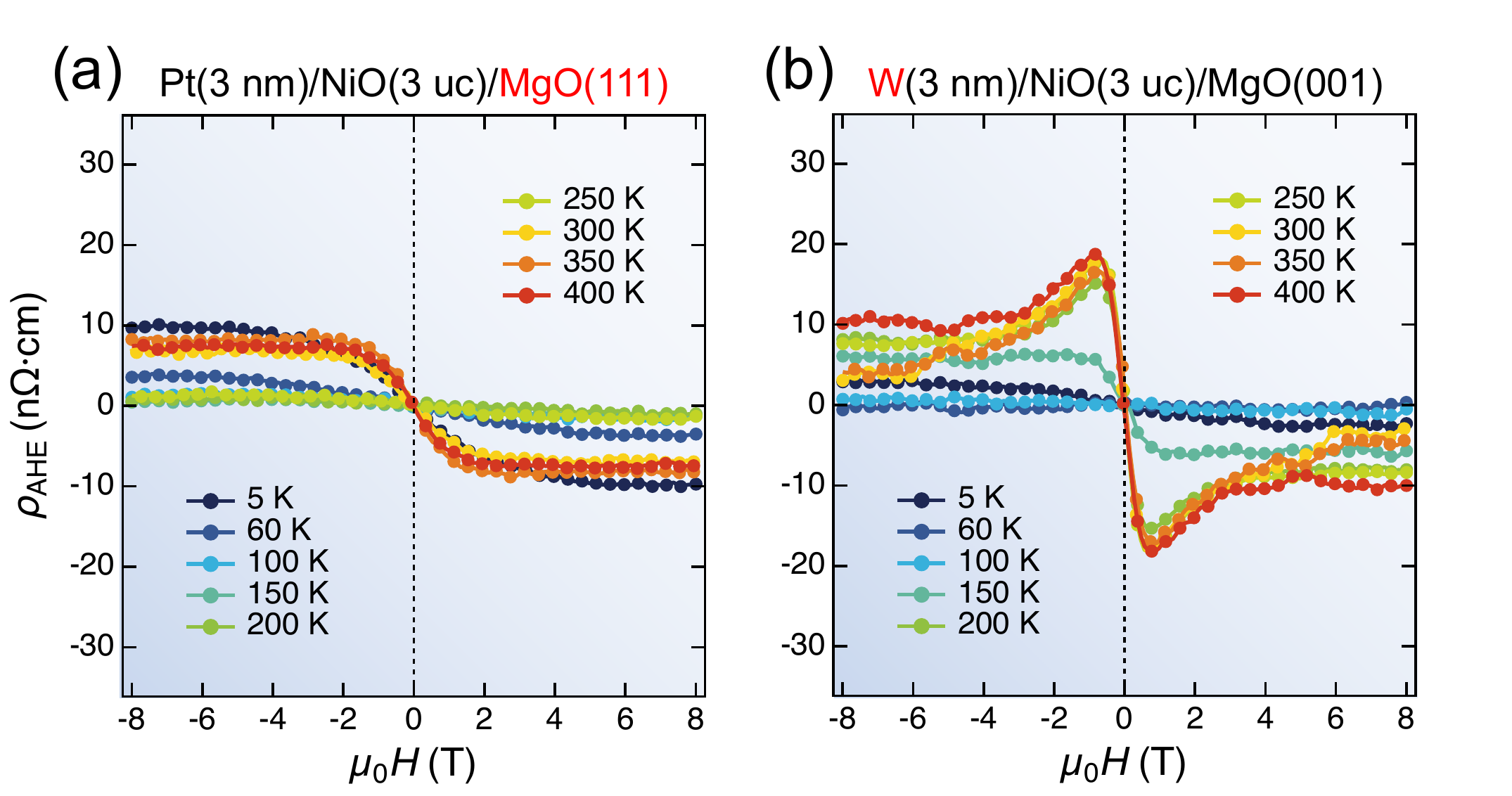}
	\caption{Insensitivity of AHE to AFMI orientation and HM type. (a-b) $\rho_{\mathrm{AHE}}$ of Pt(3 nm)/NiO(3 uc)/MgO(111) and W(3 nm)/NiO(3 uc)/MgO(001) heterostructures at various temperatures.}
	\label{Fig2}
\end{figure*}

In this study, we demonstrate that the AHE can be induced in HM/NiO heterostructures at elevated temperatures (up to 400 K) by controlling the AFM--paramagnetic (PM) (AFM--PM) transition of an ultrathin epitaxial NiO film. The value of $\rho_\mathrm{AHE}$ was approximately 40 n$\Omega$ cm, which is higher than the recorded $\rho_\mathrm{AHE}$ in HM/magnetic insulator heterostructures. The origin of the AHE in HM/AFMI heterostructures is attributed to the emergence of AFM spin textures with uncompensated topological charges during the AFM--PM phase transition. Moreover, the line shape of the AHE signal depends on the defect at the HM/AFMI interface, which can induce topological-Hall-effect-like (THE-like) signals based on the two-channel AHE scenario.

\section{Results and Discussions}

We first employed scanning transmission electron microscopy (STEM) to characterize the structural properties of the HM/NiO heterostructures with ultrathin NiO. The STEM image and selected area electron diffraction (SAED) pattern of a representative Pt(3 nm)/NiO(3 uc)/MgO(001) heterostructure are shown in Figure \ref{Fig1}a, where uc denotes unit cells (see details in Experimental Section). The NiO was fully epitaxial on the MgO(001) substrate with a rock-salt crystalline structure, whereas the Pt overlayer was polycrystalline. The field-dependent magnetization measurements showed no observable macroscopic ferromagnetism (see Figure S1,  Supporting Information). To investigate both the Pt and NiO thickness dependence of the $\rho_\mathrm{AHE}$, we performed Hall measurements at 5--400 K, with subtraction of the linear ordinary Hall signal (Complete AHE data are presented in Figure S2, Supporting Information). A low driving current density of $\sim$$10^2$ A cm$^{-2}$ was used to suppress the Joule heating. The absence of an AHE in the reference sample without the NiO layer is shown in Figure S3 (Supporting Information).

A temperature-dependent AHE was observed in the Pt(3 nm)/NiO(3 uc)/MgO(001) heterostructure (Figure \ref{Fig1}b). The AHE was only observed at high temperatures (250--400 K). This trend is consistent with that of Cr$_2$O$_3$-based HM/AFMI heterostructures.\cite{Ji2018, Cheng2019, Moriyama2020} Comprehensive contour plots of $\rho_\mathrm{AHE}$ in Pt(3 nm)/NiO($x$ uc)/MgO(001) and Pt($x$ nm)/NiO(3 uc)/MgO(001) are shown in Figure \ref{Fig1}c and Figure \ref{Fig1}d. The AHE mainly existed above 150 K, and its magnitude significantly depended on the NiO and Pt thicknesses. As the NiO thickness increased, the minimum temperature required for the appearance of $\rho_\mathrm{AHE}$ increased. For NiO thicker than 6 uc, we observed no apparent $\rho_\mathrm{AHE}$ signals within the accessible temperature (5--400 K) and field ($\pm$8 T) ranges. Figure \ref{Fig1}d shows the decay of $\rho_\mathrm{AHE}$ as Pt becomes thicker, indicating the interfacial nature of the AHE. It is noted that a local maximum of $\rho_\mathrm{AHE}$ appeared at approximately 5 K, which could be induced by the skew scattering of electrons by localized paramagnetic centers (Giovannini--Kondo model).\cite{Giovannini1973, Maryenko2017}

\begin{figure*}[th]
\centering
	\includegraphics[width=0.9\linewidth]{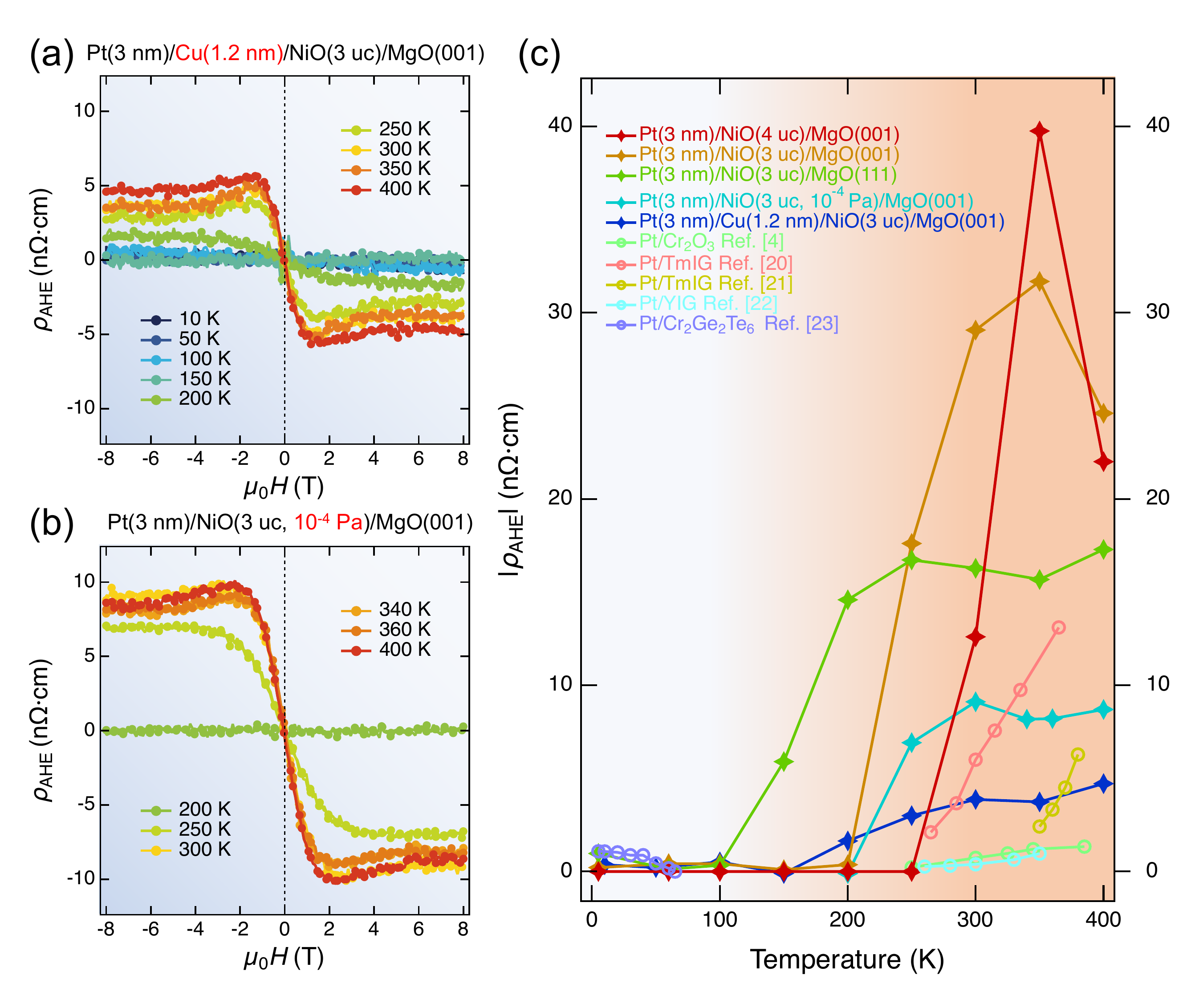}
	\caption{Control experimental results for AHE in HM/AFMI heterostructures. (a-b) $\rho_{\mathrm{AHE}}$ of Pt(3 nm)/NiO(3 uc)/MgO(111) and W(3 nm)/NiO(3 uc)/MgO(001) at various temperatures. (c) Temperature-dependent saturated $\rho_{\mathrm{AHE}}$ from our data and reported HM/magnetic insulators.\cite{Cheng2019, Ahmed2019, Shao2019, Liu2020, Lohmann2019} The four-pointed stars represent our data, and the circles represent data from references.}
	\label{Fig3}
\end{figure*}

As a $G$-type collinear AFM, the superexchange interaction between Ni ions results in antiferromagnetically stacking of ferromagnetic \{111\} planes.\cite{Baltz2018} To investigate how the crystallographic orientation of AFMI and the type of HM influence the AHE, we further measured $\rho_\mathrm{AHE}$ in the Pt(3 nm)/NiO(3 uc)/MgO(111) and W(3 nm)/NiO(3 uc)/MgO(001) heterostructures. A similar temperature-dependent AHE behavior was observed (Figure \ref{Fig2}), indicating that the AHE was insensitive to the crystal orientation of AFMI and the type of HM.  In particular, W showed the opposite sign of the spin Hall angle to Pt\cite{Ma2018}, however, the sign of AHE remained the same.\cite{Cheng2019} In addition, the line shape of $\rho_\mathrm{AHE}$ for W(3 nm)/NiO(3 uc)/MgO(001) was slightly different from that of Pt(3 nm)/NiO(3 uc)/MgO(001), which will be discussed subsequently. Additional control experiments were performed to clarify the role of the HM and concomitant spin Hall effect (SHE) and the strong spin-orbit coupling on the AHE. First, no AHE signal was observed for Ti(3 nm)/Cu(3 nm)/NiO(3 uc)/MgO(001) as expected (Figure S4, Supporting Information). Second, a weaker AHE signal was observed in Pt(3 nm)/Cu(1.2 nm)/NiO(3 uc)/MgO(001) (Figure \ref{Fig3}a), ruling out the proximity-induced-ferromagnetism in Pt. Finally, Pt/W/NiO/MgO(001) heterostructures were fabricated, and the competing of spin current\cite{Ma2018} initially decreased the magnitude of the AHE, and eventually reversed the polarity (sign) of the AHE with thickened Pt (Figure S5, Supporting Information). The experiments demonstrate that the SHE in HM plays a critical role, and the scattering of the spin current by AFMI (in a reflective manner) induces the AHE.The experimental results, along with the $\rho_\mathrm{AHE}$ values reported for HM/magnetic insulator heterostructures, are presented in Figure \ref{Fig3}c. We achieved $\rho_\mathrm{AHE}$ values up to 40 n$\Omega$ cm in the Pt/NiO heterostructures, which is higher than the recorded value in HM/magnetic insulator heterostructures.\cite{Cheng2019, Ahmed2019, Shao2019, Liu2020, Lohmann2019} Additionally, the scaling relation of the HM/AFMI heterostructure is shown in Figure S6 (Supporting Information), which is similar to that of the gating-induced ferromagnetic Pt.\cite{Shimizu2013, Liang2018} 

An additional THE-like Hall effect (the bump and dip feature added to the AHE signals) was observed in W(3 nm)/NiO (3 uc)/MgO(001) and Pt/Cu(1.2 nm)/NiO (3 uc)/MgO(001) compared to Pt/NiO (3 uc)/MgO(001) (Figure \ref{Fig2}b and \ref{Fig3}b). From the Ellingham diagram, it can be observed that the Cu/Cu$_2$O, Ni/NiO and W/WO$_3$ lines are relatively close, whereas the Pt/PtO$_2$ line is significantly higher than all the above lines.\cite{Chang2010} This implies that the oxygen can migrate from NiO to Cu and W to induce additional oxygen vacancies in NiO, resulting in weak ferromagnetism.\cite{Coey2006, Park2008, Madhu2016, Punugupati2015} To confirm this defect-induced THE-like signal, we deposited a NiO layer with a significantly lower oxygen pressure to enrich the oxygen vacancies. A THE-like signal was detected again (Figure \ref{Fig3}b), indicating that the THE-like signal was related to oxygen vacancy-induced dilute ferromagnetism. Thus, the THE-like signal can be regarded as an alternative two-channel AHE scenario engineered by the defects.\cite{Kan2018, Wu2020} One of the channels results from the AFM order, whereas the other is related to the dilute ferromagnetism, similar to the HM/FM heterostructures.\cite{Chen2013, Meyer2015}

Next, we discuss the relationship between the $T_\mathrm{N}$ of the AFMI layer and the temperature-dependent AHE behavior. The $T_\mathrm{N}$ of the ultrathin NiO is highly dependent on its thickness and boundary conditions. It has been reported that an ultrathin NiO film grown on MgO can have a significantly reduced $T_\mathrm{N}$ from its bulk value, for example, the $T_\mathrm{N}$ of NiO(3 uc)/MgO(001) was lower than 40 K, and its $T_\mathrm{N}$ could be considerably enhanced up to 390 K simply by Ag capping owing to image charge screening.\cite{Altieri2009} Therefore, the $T_\mathrm{N}$ in the same 3 uc NiO film in our Pt/NiO/MgO(001) should also be enhanced by the Pt overlayer to induce a high-temperature AHE. Note that, the short-range magnetic order above $T_\mathrm{N}$\cite{Baster2005, Rechtin1972, Hermsmeier1989, Lin2016} could also support the high-temperature AHE.

\begin{figure*}[th]
\centering
	\includegraphics[width=.7\linewidth]{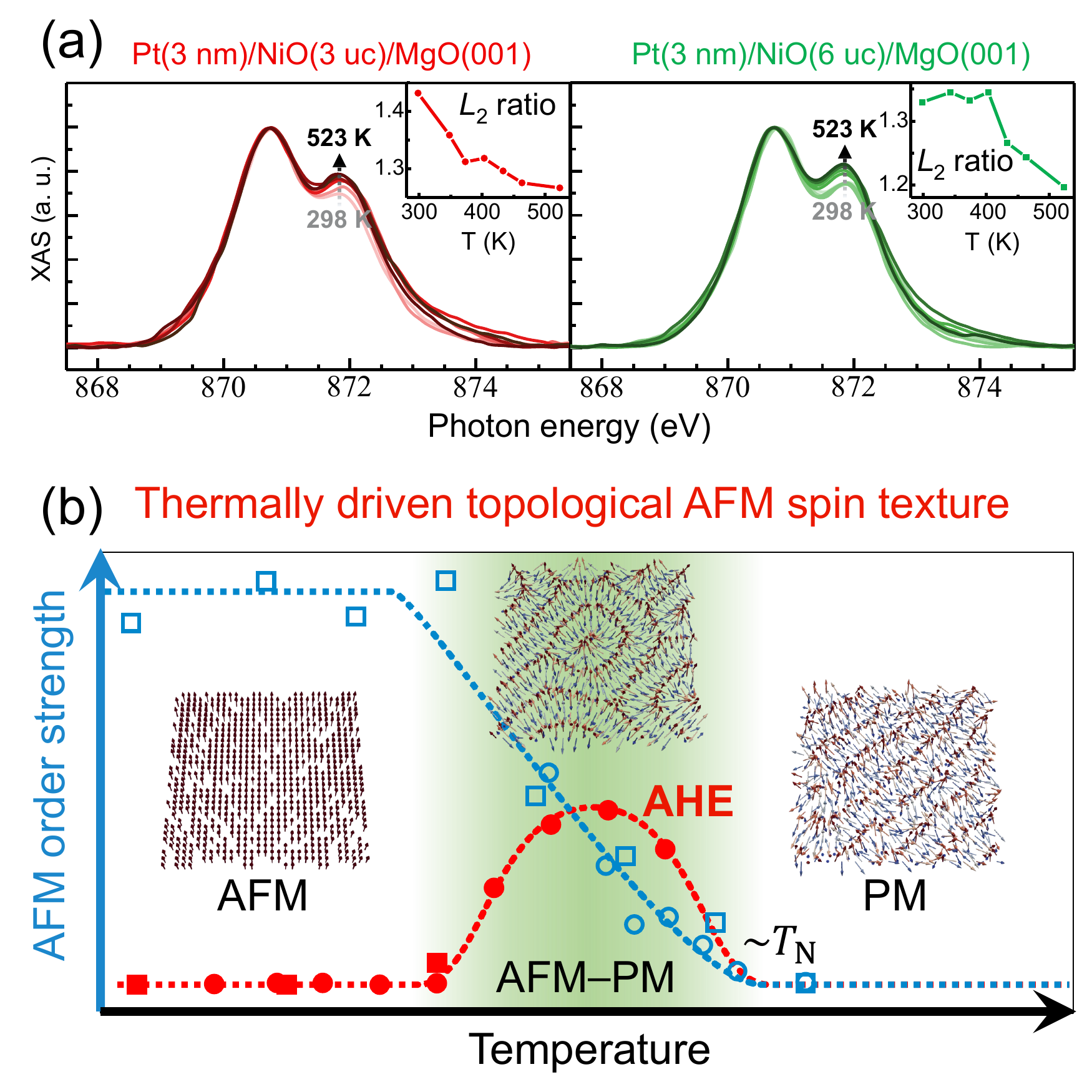}
	\caption{Correlation between AHE and AFM--PM transition. a) Temperature-dependent XAS for Pt(3 nm)/NiO(3 uc)/MgO(001) and Pt(3 nm)/NiO(6 uc)/MgO(001). b) Curves are extracted from AHE (red symbols) and XAS (blue symbols) data from above two samples (circles for the sample with 3 uc NiO, and squares for that with 6 uc NiO), normalized by the putative AFM--PM transition temperature ranges. The AFM order strength is represented by the $L_2$ ratio of XAS data. The insets show the calculated spin structures in the ultrathin NiO, represented by the distribution of N'eel vector $\mathbf{n}$, and the vector length is not proportional to the magnitude of the $\mathbf{n}$ at various temperatures.}
	\label{Fig4}
\end{figure*}

To determine the $T_\mathrm{N}$ of NiO in our Pt/NiO/MgO sample, we performed temperature-dependent X-ray absorption spectroscopy (XAS) measurements at the Ni $L_2$ edge. The decrease in the $L_2$ ratio with increasing temperature indicates an AFM--PM transition.\cite{Altieri2009, Alders1998} The continuous decrease in the Ni $L_2$ ratio of Pt(3 nm)/NiO(3 uc)/MgO(001) at 298--523 K indicates that this temperature range is within the continuous second-order AFM--PM transition range of NiO (left panel of Figure \ref{Fig4}a), and the $T_\mathrm{N}$ should be as high as 523 K and comparable to that of the bulk NiO (Here, the $T_\mathrm{N}$ is defined as the temperature of the absence of the long-range AFM order via thermal fluctuation). This indicates that Pt can impose a more pronounced effect than Ag on stabilizing the AFM order in NiO. However, for Pt(3 nm)/NiO(6 uc)/MgO(001), the Ni $L_2$ ratio started to decline above 400 K, showing that the AFM--PM transition in this sample was initiated at 400 K. W exhibited a similar effect on 3 uc NiO, that is, the AFM--PM transition covered the temperature range of 298--523K (Figure S7, Supporting Information). A comparison between the AHE and XAS data showed that the AHE coincided with that of the AFM--PM transition in the common temperature range of 300--400 K. Notably, the absence of AHE in Pt(3 nm)/NiO(6 uc)/MgO(001) below 400 K can be explained by the AFM--PM transition starting at temperatures higher than 400 K.

Based on the consistent temperature range between the AFM--PM transition and the AHE, the thermally softened AFM order during the AFM--PM transition plays a crucial role in the AHE of HM/AFMI heterostructures. In Cr$_2$O$_3$-based HM/AFMI heterostructures, the AFM--PM transition generates the temperature-dependent uncompensated magnetic moment on the HM/AFMI interface based on SH--AHE scenario, which is however questioned by X-ray magnetic dichroism measurement.\cite{Ji2018, Cheng2019, Moriyama2020} In our case, the magnetically compensated (100)-facet and uncompensated (111)-facet in NiO exhibited similar temperature-dependent AHE. Thus, the conventional SH--AHE scenario based on uncompensated magnetic moments should not account for the appearance of the AHE. 

Three mechanisms cause the conventional AHE: intrinsic contribution by the Berry phase and extrinsic contributions by skew scattering and side jump.\cite{Nagaosa2010, Feng2022} In this study, such an AHE in the heterostructure system occurs owing to the interaction between the spin current (SHE in HM) and the magnetic structure (magnetic insulator, in this case, AFMI). Recently, it has been reported that an AFM topological spin texture emerges around the phase-transition temperature of Pt/$\alpha$-Fe$_2$O$_3$ heterostructures.\cite{Jani2021} A significant Fert--Levy-type interfacial Dzyaloshinskii--Moriya interaction (DMI) exists and favors noncollinear spin textures, considering the space-inversion symmetry breaking on the interface.\cite{Fert1980} Thus, the AFM topological spin textures could exist in HM/AFMI heterostructures at room temperature, responsible for the AHE by scattering spin-polarized electrons with an effective field $\langle b_\mathrm{z}\rangle = \sqrt{3}\phi_\mathrm{0}/(2\lambda^{2})$ for per unit topological charge, where $\langle b_\mathrm{z}\rangle$ is the out-of-plane effective field, $\phi_\mathrm{0}$ is the unit flux and $\lambda$ is the size of spin texture.\cite{Nakatsuji2015, Nagaosa2013}

We performed atomistic spin dynamics simulations to investigate the formation of AFM topological spin textures. We found that a noncollinear AFM spin texture could emerge during the AFM--PM transition (Figure \ref{Fig4}b) owing to the interaction among the collinear AFM exchange coupling, DMI, thermal field, and external magnetic field (see the details in Experimental Section). In the intermediate-temperature region, the thermal energy $k_\mathrm{B}T$ is sufficiently high to destabilize the otherwise collinear AFM order (for example, the spin structure in Figure \ref{Fig4}b with $k_\mathrm{B}T = 0$), such that interfacial DMI can induce the noncollinear AFM spin texture. At high temperatures, the thermal perturbation dominates, yielding a PM-like spin state. The curves in Figure \ref{Fig4}b are guides for the eye to show the coincident temperature range between the AFM--PM transition and the AHE occurrence (the putative AFM--PM transition ranges for sample Pt(3 nm)/NiO(3 uc)/MgO(001) and Pt(3 nm)/NiO(6 uc)/MgO(001) are approximately 200--523 K and 400--523 K, respectively.). 

The topological charges associated with the two spin sublattices are denoted as $Q_\mathrm{A}$ and $Q_\mathrm{B}$, respectively, and can be calculated from the spatial distribution of the N\'eel vector (see Experimental Section and Figure S8, Supporting Information). For a canonical AFM skyrmion, the local effective fields induced by $Q_\mathrm{A}$ and $Q_\mathrm{B}$ have the same magnitude but opposite signs because $Q_\mathrm{A} + Q_\mathrm{B} = 0$ and $|Q_\mathrm{A}| = |Q_\mathrm{B}| = 1$ leading to the generation of a zero net effective field.\cite{Nagaosa2013} Therefore, an AFM skyrmion cannot cause an AHE because the time-reverse symmetry is preserved. However, applying a large out-of-plane magnetic field can break the time-reverse symmetry as the two spin sublattices are canted (Figure S9, Supporting Information), yielding a non-zero net topological charge ($Q_\mathrm{A} + Q_\mathrm{B} \neq 0$) and hence, an AHE.

\section{Conclusion}

In this paper, we reported an unconventional high-temperature AHE insensitive to the selection of the HM and AFMI in HM/AFMI heterostructures, which was suppressed and even eliminated in a low-temperature region. This is explained by an extended SMR model that includes the AFM spin texture. The emergence of the AFM spin texture (for example, skyrmion or meron) is considered to be stabilized by the DMI at the HM/AFMI interface, and can only occur at an intermediate temperature during the AFM--PM phase transition, as demonstrated via micromagnetic simulations. The THE-like Hall effect can be triggered by controlling the fabrication conditions of the samples, which can be regarded as an additional AHE channel induced by defect-induced weak FM. The room-temperature AHE in AFMI heterostructures demonstrates the strong interaction between the AFM order and spin current, which aids in developing AFM-based spintronics devices.

\section{Experimental Section} \label{sec3}

\noindent
\textbf{Sample fabrication}\\
The epitaxial NiO thin films were fabricated using pulsed laser deposition (PLD). The growth conditions were as follows: a growth temperature of 650 \degree C (measured using a pyrometer), oxygen background pressure of 50 mTorr, an excimer laser with a wavelength of 248 nm, a repetition rate of 3 Hz, an energy density of 1.4 J cm$^{-2}$, and target-substrate distance of 5.5 cm. After deposition and cooling down to room temperature, the NiO films were immediately transferred to a magnetron sputtering chamber (AJA International, Inc.) with a background pressure better than $2 \times 10^{-8}$ torr, and heated to 150 \degree C for 15 minutes, to prevent possible surface gas absorption. Then, DC sputtering was then employed to deposit the Pt layer at room temperature with a power of 30 W, background Ar of 3 mTorr, resulting in a deposition rate of 2.6 nm/min. The W and Cu were deposited under the same conditions at deposition rates of 1.3 nm/min and 3.6 nm/min, respectively.

\bigskip

\noindent
\textbf{Structural, electrical and magnetic properties characterization}\\
The structural properties were characterized using a high-resolution X-ray diffractometer (XRD, Malvern Panalytical). The magnetic hysteresis loop was measured using a superconducting quantum interference device (MPMS, Quantum Design). The transport properties were measured using the van der Pauw method in a Physical Property Measurement System (PPMS, Quantum Design DynaCool system). The measured Hall signals were firstly treated using an asymmetry procedure, and the ordinary Hall signals were then subtracted. Cross-sectional high-resolution transmission electron microscopy samples were prepared using a focused ion beam (FIB, Zeiss Auriga) with Ga$^{+}$ ions.

\bigskip

\noindent
\textbf{X-ray absorption spectroscopy}\\
Element-specific X-ray absorption spectroscopy (XAS) was measured using total electron yield (TEY) at Beamline BL02B02 of the Shanghai Synchrotron Radiation Facility (SSRF) and Beamline 4B9B of the Beijing Synchrotron Radiation Facility (BSRF). To evaluate the magnetic properties of these NiO thin films, we utilized the magnetic linear dichroism (MLD) effect in the Ni $L_2$ XAS. The normal incidence of X-rays resulted in the electric field component of X-rays being parallel to the sample plane. The sample temperature was controlled by laser heating the sample holder. A linear background connecting raw data points at 870.8 eV and 871.8 eV was first subtracted from the raw data, and the spectra were normalized from 0 to 1. The $L_2$ ratio was defined as the normalized peak intensity at 871.8 eV.

\bigskip

\noindent
\textbf{Atomistic spin dynamics simulations}\\
Micromagnetic simulations were performed to simulate the spin structure in the AFM NiO with interfacial DMI. The total Hamiltonian of the AFM thin-film $\mathcal{H}$ included the Hamiltonians of the exchange interaction $\mathcal{H}_\mathrm{exch}$, uniaxial anisotropy $\mathcal{H}_\mathrm{anis}$, external fields $\mathcal{H}_\mathrm{ext}$, and the Hamiltonian of the interfacial Dyzaloshinskii–Moriya interaction (DMI) $\mathcal{H}_\mathrm{DMI}$, expressed as follows:
\begin{equation}
\begin{aligned}[b]
  \mathcal{H} = & \mathcal{H}_\mathrm{exch} + \mathcal{H}_\mathrm{anis} + \mathcal{H}_\mathrm{ext} + \mathcal{H}_\mathrm{DMI} \\
= & -\sum_{<k, l>} J \mathbf{S}_{k} \cdot \mathbf{S}_{l}-\sum_{k} K_{a}\left(\mathbf{S}_{k} \cdot \mathbf{n}_{e}\right)^{2} \\ &+ \sum_{<k, l>} \mathbf{D}_{k l} \cdot\left(\mathbf{S}_{k} \times \mathbf{S}_{l}\right) \ ,  
\end{aligned}
\end{equation}
where $\mathbf{S}_{k}$ and $\mathbf{S}_{l}$ are the orientation vectors of the local spin and neighboring spins, respectively; $J$ is the exchange constant, $K_{a}$ is the uniaxial anisotropy constant, $\mathbf{n}_{e}$ is the uniaxial easy axis, and $\mathbf{D}_{k l}$ is the DMI constant. 

In the simulation, a two-dimensional system of 64 $\times$ 64 grids was considered in which each grid with a size of 1 nm $\times$ 1 nm, contained two neighboring spins. By linearly combining $\mathbf{S}_{m}$ and $\mathbf{S}_{n}$ inside the same gird, the net magnetization vector $\mathbf{m} = (\mathbf{S}_{m} + \mathbf{S}_{n})/2$ and the N\'eel vector $\mathbf{n} = (\mathbf{S}_{m} - \mathbf{S}_{n})/2$ were defined. The expression of the free energy was derived by substituting $\mathbf{S}_{m}$ and $\mathbf{S}_{n}$ with $\mathbf{m}$ and $\mathbf{n}$, respectively, in the total Hamiltonian. Under the continuum approximation\cite{Li20202}, the total free energy is expressed as follows: 
\begin{equation}
\begin{aligned}[b]
E_{\mathrm{tot}} =&\int d V\bigg\{A^{*}\left[\left(\partial_{x} \mathbf{n}\right)^{2}+\left(\partial_{y} \mathbf{n}\right)^{2}\right]-K\left(\mathbf{n} \cdot \mathbf{n}_{e}\right)^{2} \\
&+D\left[n_{z}(\nabla \cdot \mathbf{n})-(\mathbf{n} \cdot \nabla) n_{z}\right]-\mathbf{H}_{\mathrm{ext}} \cdot \mathbf{n}\bigg\} \ ,
\label{eq2}
\end{aligned}
\end{equation}
where $A^{*} = -2.42$ pJ m$^{-1}$, is the continuum-scale exchange constant, which was converted from the AFM exchange interaction $J = -19.01$ meV\cite{chatterji2009}. We assumed the existence of a perpendicular magnetic anisotropy (i.e., $\mathbf{n}_e//z$) in the ultrathin (3 u.c.) NiO owing to the interfacial symmetry breaking, with a continuum anisotropy constant $K=1.5 \times 10^5$  J m$^{-3}$(which is a typical number utilized in existing micromagnetic simulation investigations of AFM skyrmions)\cite{Liang2021}, $D^* = -3.5 \times 10^{-3}$ J m$^{-2}$ is the continuum DMI constant, which is a value calculated from the first-principles in a similar NiO/Au system.\cite{Akanda2020}

The free energy expression of the N\'eel vector $\mathbf{n}$ in Eq. (\ref{eq2}) is the same as that of the magnetization in the ferromagnetic materials. Therefore, we performed micromagnetic simulations using the open-source software MuMax3\cite{Vansteenkiste2014} to calculate the equilibrium configuration of the N\'eel vector. The evolution of $\mathbf{n}$ was determined by solving the Landau--Lifshitz--Gilbert (LLG) equation,
\begin{equation}
\frac{\partial \mathbf{n}}{\partial t}=-\frac{\gamma_{0}}{1+\alpha^{2}}\left(\mathbf{n} \times \mathbf{H}_{\mathrm{eff}}+\alpha \mathbf{n} \times \mathbf{n} \times \mathbf{H}_{\mathrm{eff}}\right) \ ,
\end{equation}
where $\alpha = 0.01$ is the Gilbert damping coefficient\cite{Akanda2020} and $\gamma_0$ is the gyromagnetic ratio; $\mathbf{H}_{\mathrm{eff}} = - \frac{1}{\mu_0 M_s}\frac{\delta E_{\mathrm{tot}}}{\delta \mathbf{n}}$ is the effective field, where $\mu_0$ is the vacuum permeability, amd $M_s = 6.2 \times 10^5$ A m$^{-1}$ is the saturation magnetization of NiO\cite{Akanda2020}. 

The influence of thermal fluctuations was modeled by adding a thermal fluctuation field $\mathbf{H}_{\mathrm{therm}}$ to $\mathbf{H}_{\mathrm{eff}}$, expressed by $\mathbf{H}_{\text {therm }}=\bm{\upeta} \sqrt{\frac{2 \alpha k_{B} T}{\mu_{0} M_{s} \gamma_{0} \Delta V \Delta t}}$, where $k_{B} T$ is the Boltzmann constant, T is the temperature, $\Delta V$ is the volume of each simulation cell, and $\Delta t$ is the time interval in a real unit. $\bm{\upeta} = \bm{\upeta} (\mathbf{r}, t) = (\eta_x, \eta_y, \eta_z)$ is a white-noise distributed stochastic vector uncorrelated in space and time. The temporal mean value of $\eta_i (i = x, y, z)$ was zero. The LLG equation was solved using fourth-order Runge--Kutta methods, with a discretized time interval of 20 fs. A uniform distribution of $\mathbf{n}$ in the $z$-direction was used as the initial state for the low-temperature region ($<$ 8 K), and an artificially generated single AFM skyrmion was used for the higher-temperature region ($>$ 8 K). The equilibrium distribution of $\mathbf{n}$ was assumed to be reached when the total free energy density $E_\text{tot}$ no longer changes significantly with time.

The topological charge of the two sublattices of $\mathbf{n}$ in the equilibrium state, $Q_\mathrm{A}$ and $Q_\mathrm{B}$ were calculated as follows to quantify the topological state of the AFM thin film\cite{Zhang2016}:
\begin{equation}
\begin{aligned}
&Q_\mathrm{A}=\frac{1}{8 \pi} \int_{\text {site A}} q_{ijk} \ ; \\
&Q_\mathrm{B}=\frac{1}{8 \pi} \int_{\text {site B}} q_{ijk} \ ,
\end{aligned}
\end{equation}
where $q_{ijk}$ is calculated for each unique triangle with grids $i$, $j$, and $k$ as vertices using Equation (\ref{eq5})
\begin{equation}
\tan(\frac{q_{ijk}}{2}) = \frac{\mathbf{n}_i \cdot (\mathbf{n}_j \times \mathbf{n}_k)}{1+\mathbf{n}_i \cdot \mathbf{n}_j + \mathbf{n}_i \cdot \mathbf{n}_k + \mathbf{n}_j \cdot \mathbf{n}_j} \ .
\label{eq5}
\end{equation}
The equilibrium state of $\mathbf{n}$ was determined based on the values of $Q_\mathrm{A}$ and $Q_\mathrm{B}$. If $Q_\mathrm{A} + Q_\mathrm{B} = 0$, the distribution of $\mathbf{n}$ is identified as a collinear distribution, otherwise, it is a topological AFM spin texture. If $Q_\mathrm{A} \neq Q_\mathrm{B}$, there is no evident ordered spin texture\cite{Li20202}.

\bigskip
\noindent
\textbf{Acknowledgements} \par 
Y. L. and L. W. contributed equally to this study. We acknowledge the insightful discussions with Shuai Dong and Jiahao Han. We thank the staff at Beamline 4B9B of the BSRF and Beamline 02B02 of the SSRF for fruitful discussions and experimental assistance. L. W. acknowledges the support from the Natural Science Foundation of China (Grant No. 52102131) and Yunnan Fundamental Research Projects (Grant Nos. 202101BE070001-012 and 202201AT070171). C.-W. N and Y.-H. L were supported by the Basic Science Center Project of the National Natural Science Foundation of China (NSFC) (Grant No. 51788104). The work conducted at the University of Wisconsin-Madison was supported by the National Science Foundation (NSF) under the Award CBET-2006028. The atomistic spin dynamics simulations were performed using Bridges at the Pittsburgh Supercomputing Center under Allocation TGDMR180076, which is a part of the Extreme Science and Engineering Discovery Environment (XSEDE) and supported by NSF Grant ACI-154856. Y.Z. would like to thank the support from the Natural Science Foundation of China (Grant No.52002370).

\bibliographystyle{apsrev4-2}
\bibliography{ref}
\end{document}